# Two Relaxation Time Lattice Boltzmann Method Coupled to Fast Fourier Transform Poisson Solver: Application to Electroconvective Flow

Yifei Guan and Igor Novosselov [§]

*Department of Mechanical Engineering, University of Washington, Seattle, U.S.A. 98195*
December 2018

## I. ABSTRACT

Electroconvective flow between two infinitely long parallel electrodes is investigated via a multiphysics computational model. The model solves for spatiotemporal flow properties using two-relaxation-time Lattice Boltzmann Method for fluid and charge transport coupled to Fast Fourier Transport Poisson solver for the electric potential. The segregated model agrees with the previous analytical and numerical results providing a robust approach for modeling electrohydrodynamic flows.

## II. KEYWORDS

Two-Relaxation-Time Lattice Boltzmann Method; Fast Poisson Solver; Electroconvection Stability

## III. INTRODUCTION

Electro-hydrodynamics (EHD) studies the interaction of the fluid with electric field [1]. As a subset of EHD, electroconvection (EC) is a phenomenon where convective transport is induced by unipolar discharge into a dielectric fluid [2-21]. Felici first performed a stability analysis of EC using a non-linear hydraulic model [22,23]. Linear stability analysis was investigated by Schneider & Watson [24,25] and Atten & Moreau [26], who showed that, in the weak-injection limit, $C \ll 1$, where $C$ is the charge injection, the flow stability is determined by the criterion $T_c C^2$, where $T_c$ is the linear stability threshold for the electric Rayleigh number ($T$) – a ratio between electric force to the viscous force. In the space-charge-limited (SCL) injection ($C \rightarrow \infty$), the flow stability is determined by $T_c$. Experimental observations of Lacroix et al. [27] and Atten et al. [28] showed that, in the SLC limit, $T_c = 100$ [28], while linear stability analysis suggests $T_c = 160.45$ for the same conditions [26]. Authors suggest that the discrepancy is due to the omission of the charge diffusion term in the analysis [29]. The effect of charge diffusion was investigated by Zhang et al., who performed a linear stability analysis [12] followed by a non-linear analysis using the multiscale method [17]. Zhang et al. found that the charge diffusion has a non-negligible effect on $T_c$, but their analysis could not bridge the discrepancy between the experimental and theoretical values.

To gain insight into the flow-charge interaction, the EC problem has been investigated by numerical methods. Castellanos and Atten used a finite difference model, concluding that large numerical diffusivity can contaminate the model [3]. Other numerical models used to study EC phenomena include the particle-in-cell method [30], finite volume method with flux-corrected transport [31], total variation diminishing scheme [5,8,14-16], and the method of characteristic [4]. Recently, Luo et al. showed that a unified Lattice Boltzmann model (LBM) matches the linear and finite amplitude stability criteria of the subcritical bifurcation in EC flow [18-21] for both 2D and 3D flow scenarios. This unified LBM transforms the elliptic Poisson equation to a parabolic reaction-diffusion equation and introduces artificial coefficients to control the evolution of the electric potential, which may result in ambiguity in the numerical results.

In this paper, we demonstrate an alternative approach to modeling EC flow; our segregated solver combines (i) a two-relaxation-time (TRT) LBM for modeling fluid transport and charged species, and (ii) a Fast Fourier Transform (FFT) Poisson approach to directly solve for the electric field. The TRT model introduces two relaxation parameters aiding the numerical algorithm stability without sacrificing computational efficiency.

[§] ivn@uw.edu

## IV. NUMERICAL METHOD

### 1. Governing Equations

The governing equations for EHD flow include the Navier-Stokes equations (NSE), with the addition of an electric forcing term $\mathbf{F_e} = -\rho_c \nabla \varphi$ to the momentum equation, the charge transport equation, and the Poisson equation for electric potential.

$$\nabla \cdot \mathbf{u} = 0, \tag{1}$$

$$\rho \frac{D\mathbf{u}}{Dt} = -\nabla P + \mu \nabla^2 \mathbf{u} - \rho_c \nabla \varphi, \tag{2}$$

$$\frac{\partial \rho_c}{\partial t} + \nabla \cdot \left[ (\mathbf{u} - \mu_b \nabla \varphi) \rho_c - D_c \nabla \rho_c \right] = 0, \tag{3}$$

$$\nabla^2 \varphi = -\frac{\rho_c}{\varepsilon}, \tag{4}$$

where $\rho$ is the density, $\mu$ - dynamic viscosity, $\mathbf{u} = (u_x, u_y)$ - velocity vector field, $P$ - static pressure, $\mu_b$ - ion mobility, $D_c$ - ion diffusivity, $\rho_c$ - charge density, $\varepsilon$ - electric permittivity, $\varphi$ - electric potential. The electric force provides a source term in the momentum equation (Eq. 2).

### 2. Lattice Boltzmann Method

The TRT-LBM is applied to NSE (Eq.1-2) and the transport equation for charge density (Eq. 3). The mesoscopic solutions of the LBM yield a discrete distribution function of velocity $f_i(\mathbf{x},t)$ and charge density $g_i(\mathbf{x},t)$. The values of $\rho$, $\rho_c$, and momentum density $\rho\mathbf{u}$ can be evaluated by weighted sums.

$$\rho(\mathbf{x},t) = \sum_i f_i(\mathbf{x},t), \tag{5}$$

$$\rho_c(\mathbf{x},t) = \sum_i g_i(\mathbf{x},t), \tag{6}$$

$$\rho\mathbf{u}(\mathbf{x},t) = \sum_i \mathbf{c}_i f_i(\mathbf{x},t) + \frac{\mathbf{F}_e \Delta t}{2} = \sum_i \mathbf{c}_i f_i(\mathbf{x},t) - \frac{\rho_c \nabla \varphi \Delta t}{2}, \tag{7}$$

The discrete normalized velocity, $\mathbf{c}_i = (c_{ix}, c_{iy})$ at position - $\mathbf{x}$ and time - $t$ depends on a specific discretization scheme; here, we use the D2Q9 model (two spatial dimensions and nine discrete velocities). The spatial discretization is uniform ($\Delta x = \Delta y$), and the temporal discretization - $\Delta t$. The $\mathbf{c}_i$ parameters ($i$=0~8) are shown in the Supplementary Material.

The Lattice Boltzmann Equations (LBEs) for flow field and charge density are:

$$f_i(\mathbf{x}+\mathbf{c}_i\Delta t, t+\Delta t) = f_i(\mathbf{x},t) \underbrace{-\Delta t \left[ \omega^+ \left( f_i^+ - f_i^{eq+} \right) + \omega^- \left( f_i^- - f_i^{eq-} \right) \right]}_{\text{TRT collision operator}} + \underbrace{\Delta t \left[ \left(1 - \frac{\omega^+ \Delta t}{2}\right) F_i^+(\mathbf{x},t) + \left(1 - \frac{\omega^- \Delta t}{2}\right) F_i^-(\mathbf{x},t) \right]}_{\text{TRT source operator}}, \tag{8}$$

$$g_i(\mathbf{x}+\mathbf{c}_i\Delta t, t+\Delta t) = g_i(\mathbf{x},t) \underbrace{-\Delta t \left[ \omega_g^+ \left( g_i^+ - g_i^{eq+} \right) + \omega_g^- \left( g_i^- - g_i^{eq-} \right) \right]}_{\text{TRT collision operator}}, \tag{9}$$

$f_i^{eq}$ and $g_i^{eq}$ are the equilibrium distributions for flow field and charges respectively, which are given by

$$f_i^{eq}(\mathbf{x},t) = w_i \rho \left( 1 + \frac{\mathbf{u} \cdot \mathbf{c}_i}{c_s^2} + \frac{(\mathbf{u} \cdot \mathbf{c}_i)^2}{2c_s^4} - \frac{\mathbf{u} \cdot \mathbf{u}}{2c_s^2} \right), \tag{10}$$

$$g_i^{eq}(\mathbf{x},t) = w_i \rho_c \left( 1 + \frac{(\mathbf{u} - \mu_b \nabla \varphi) \cdot \mathbf{c}_i}{c_s^2} + \frac{[(\mathbf{u} - \mu_b \nabla \varphi) \cdot \mathbf{c}_i]^2}{2c_s^4} - \frac{(\mathbf{u} - \mu_b \nabla \varphi) \cdot (\mathbf{u} - \mu_b \nabla \varphi)}{2c_s^2} \right), \tag{11}$$

$F_i$ is the forcing term accounting for the electric force

$$F_i = w_i \left( \frac{\mathbf{c}_i - \mathbf{u}}{c_s^2} + \frac{(\mathbf{c}_i \cdot \mathbf{u}) \mathbf{c}_i}{c_s^4} \right) \cdot \mathbf{F}_e, \tag{12}$$

where $c_s^2 = P/\rho = (1/3)(\Delta x / \Delta t)^2$ is the speed of sound [32]; $\tau^\pm = 1/\omega^\pm$ and $\tau_g^\pm = 1/\omega_g^\pm$ are the times at which the distribution functions relax to equilibrium; and $w_i$ is the weight for the velocity component $\mathbf{c}_i$.

The TRT approach has been previously used to model the collision operators [33-37] and the momentum source operator [38,39]. Alternatively, to SRT operators, TRT is more robust for EC problems, as it provides additional relaxation parameter, improving the numerical stability [32]. The terms specified in the TRT collision operators and the source operator are

$$f_i^+ = \frac{f_i + f_{\bar{i}}}{2}, \ f_i^- = \frac{f_i - f_{\bar{i}}}{2}, \ f_i^{eq+} = \frac{f_i^{eq} + f_{\bar{i}}^{eq}}{2}, \ f_i^{eq-} = \frac{f_i^{eq} - f_{\bar{i}}^{eq}}{2} \tag{13}$$

$$g_i^+ = \frac{g_i + g_{\bar{i}}}{2}, \ g_i^- = \frac{g_i - g_{\bar{i}}}{2}, \ g_i^{eq+} = \frac{g_i^{eq} + g_{\bar{i}}^{eq}}{2}, \ g_i^{eq-} = \frac{g_i^{eq} - g_{\bar{i}}^{eq}}{2} \tag{14}$$

$$F_i^\pm = \frac{F_i \pm F_{\bar{i}}}{2} \tag{15}$$

Subscript $\bar{i}$ denotes the velocity component opposite to $i$, such that $\mathbf{c}_i = -\mathbf{c}_{\bar{i}}$. Relaxation parameters $\omega^+, \omega_g^+$ are determined by

$$v = c_s^2 \left( \frac{1}{\omega^+} - \frac{\Delta t}{2} \right), \ D_c = c_s^2 \left( \frac{1}{\omega_g^+} - \frac{\Delta t}{2} \right). \tag{16}$$

$\omega^-$ and $\omega_g^-$ need to satisfy

$$\Lambda = \left( \frac{1}{\omega^+ \Delta t} - \frac{1}{2} \right)\left( \frac{1}{\omega^- \Delta t} - \frac{1}{2} \right), \ \Lambda_g = \left( \frac{1}{\omega_g^+ \Delta t} - \frac{1}{2} \right)\left( \frac{1}{\omega_g^- \Delta t} - \frac{1}{2} \right), \tag{17}$$

where $\Lambda$ and $\Lambda_g$ are free factors used to control the algorithm stability [32]. Here, $\Lambda = 1/12$ and $\Lambda_g = 10^{-6}$; the large difference accounts for the mismatch between the neutral molecule and charge diffusivity.

### 3. Fast Poisson Solver

The Poisson equation (Eq. 4) is solved by a fast Poisson solver using a 2D FFT algorithm. The discretized grid function can be written as:

$$[D_x^2 + D_y^2]\varphi_{x,y} = s_{x,y} \tag{18}$$

where $D_x^2$ and $D_y^2$ are 2nd order derivatives operators in x - y coordinates; $s_{x,y}$ - source term representing space charge effect. Fourier spectral method is used in the x-direction and 2nd order finite difference scheme in the y-direction. In x-direction, the FFT algorithm is used to implement the standard Discrete Fourier transform (DFT).

$$DFT_x\left[\varphi_{x,y}\right] = \sum_{x=1}^{NX} \varphi_{x,y} \exp\left[-i\frac{2\pi(k_x-1)}{NX}(x-1)\right], \quad 1 \le k_x \le NX. \tag{19}$$

where $k_x$ is the wavenumber and $NX$ is the number of grid points in the x-direction. The 2nd derivative in the x-direction can be calculated in the Fourier domain

$$DFT_x\left[D_x^2 \varphi_{x,y}\right] = DFT_x\left[\frac{\partial^2 \varphi_{x,y}}{\partial x^2}\right] = -k_x^2 DFT_x\left[\varphi_{x,y}\right]. \tag{20}$$

Fourier transform in the y-direction uses an odd extension of the domain to satisfy the Dirichlet boundary conditions.

$$\varphi_{x,y_{ext}}^{ext} = \left[0, \varphi_{x,1}, \varphi_{x,2}, \cdots, \varphi_{x,Ny}, 0, -\varphi_{x,Ny}, \cdots, -\varphi_{x,1}\right], \tag{21}$$

$$s_{x,y_{ext}}^{ext} = \left[0, s_{x,1}, s_{x,2}, \cdots, s_{x,Ny}, 0, -s_{x,Ny}, \cdots, -s_{x,1}\right], \tag{22}$$

where $NY$ is the number of grid points in the y-direction. The size of the extended matrices is $NX \times NE$, where $NE = 2NY + 2$; the $y_{ext}$ is the extended y indices, ranging from 1 to $NE$.

From the definition of DFT (Eq.19), we have:

$$DFT_y\left[\varphi_{x,y+1}^s\right] = \exp(ik_y \Delta y) DFT_y\left[\varphi_{x,y}\right], \tag{23}$$

where $\varphi^s$ is a periodically shifted vector by $\Delta y$ of $\varphi$ in y-direction. Applying a central differencing operator in $y_{ext}$ direction gives:

$$DFT_{y_{ext}}\left[D_{y_{ext}}^2 \varphi_{x,y_{ext}}^{ext}\right] = \frac{\exp(ik_{y_{ext}}\Delta y) + \exp(-ik_{y_{ext}}\Delta y) - 2}{\Delta y^2} DFT_{y_{ext}}\left[\varphi_{x,y_{ext}}^{ext}\right] = \frac{-4\sin^2(k_{y_{ext}}\Delta y/2)}{\Delta y^2} DFT_{y_{ext}}\left[\varphi_{x,y_{ext}}^{ext}\right]. \tag{24}$$

Therefore, the Fourier transform of Eq.18 is

$$-\left(k_x^2 + \frac{4\sin^2(k_{y_{ext}}\Delta y/2)}{\Delta y^2}\right) DFT_{x,y_{ext}}\left[\varphi_{k_x,k_{y_{ext}}}^{ext}\right] = DFT_{x,y_{ext}}\left[s_{k_x,k_{y_{ext}}}^{ext}\right] \tag{25}$$

The Inverse Fast Fourier Transform (IFFT) algorithm transforms $DFT_{x,y_{ext}}\left[\varphi_{k_x,k_{y_{ext}}}^{ext}\right]$ into the spatial domain. Then, the electric potential in the original domain is obtained by retaining the first half ($1 \le y \le Ny$) of the extended solution matrix.

## 4. Boundary Conditions and Method Implementation

The numerical method is implemented in C++ using CUDA GPU computing. The number of threads in the x-direction in each GPU block is equal to $NX$; the number of GPU blocks in the y-direction is equal to $NY$. FFT and IFFT operations are performed using the cuFFT library. All variables are computed with double precision to reduce truncation errors. The numerical method was shown to be 2nd order accurate in space. Error analysis is provided in supplementary materials. To reduce computational cost while maintaining accuracy, the grid of $NX = 122$, $NY = 100$ is used throughout this work. Macroscopic and mesoscopic boundary conditions are specified in **Table I**.

Table I. Boundary conditions used in the numerical simulations.

| Boundary | Macro-variables Conditions | Meso-variables Conditions |
| --- | --- | --- |
| x direction boundaries | Periodic | Periodic |
| Upper wall | $\mathbf{u}=0$, $\varphi=0$ and $\nabla \rho_c = 0$ | Bounce-back for $f_i$ [32] |
| | | Bounce-back for $g_i$ [32] |
| Lower wall | $\mathbf{u}=0$, $\varphi=\varphi_0$ and $\rho_c=\rho_0$ | Bounce-back for $f_i$ |
| | | $\dfrac{\partial g_i}{\partial z}=0$ |

## V. RESULT AND DISCUSSION

### 1. Non-dimensional Analysis and Solution for Hydrostatic State

Governing equations yield four non-dimensional parameters that describe the system's state [12,17-21].

$$M = \frac{(\varepsilon/\rho)^{1/2}}{\mu_b},\quad T=\frac{\varepsilon\Delta\varphi_0}{\mu\mu_b},\quad C=\frac{\rho_0 H^2}{\varepsilon\Delta\varphi_0},\quad Fe=\frac{\mu_b\Delta\varphi_0}{D_e}, \qquad (26)$$

where $H$ is the distance between the electrodes (distance between the two infinite plates), $\rho_0$ is the injected charge density at the anode, and $\Delta\varphi_0$ is the voltage difference applied to the electrodes. The physical interpretation of these parameters are as follows: $M$ - ratio between hydrodynamic mobility and the ionic mobility; $T$ - a ratio between electric force to the viscous force; $C$ - charge injection level; and $Fe$ - reciprocal of the charge diffusivity coefficient [12,17].

FIG. 1 shows that our hydrostatic solutions for electric field and charge density agree well with the model of Luo et al. [18,19] and the analytical solution [30,40]. The analytical solution is based on a reduced set of equations for the electric field in one-dimensional coordinates.

$$\rho_c = \rho_a (y+y_a)^{-1/2}, \qquad (27)$$

$$E_y = \frac{2\rho_a}{\varepsilon}(y+y_a)^{1/2}, \qquad (28)$$

where $\rho_a$ and $y_a$ are parameters that depend on the boundary conditions and geometry. For the hydrostatic state, parameter $C$ dominates the system [12,17].

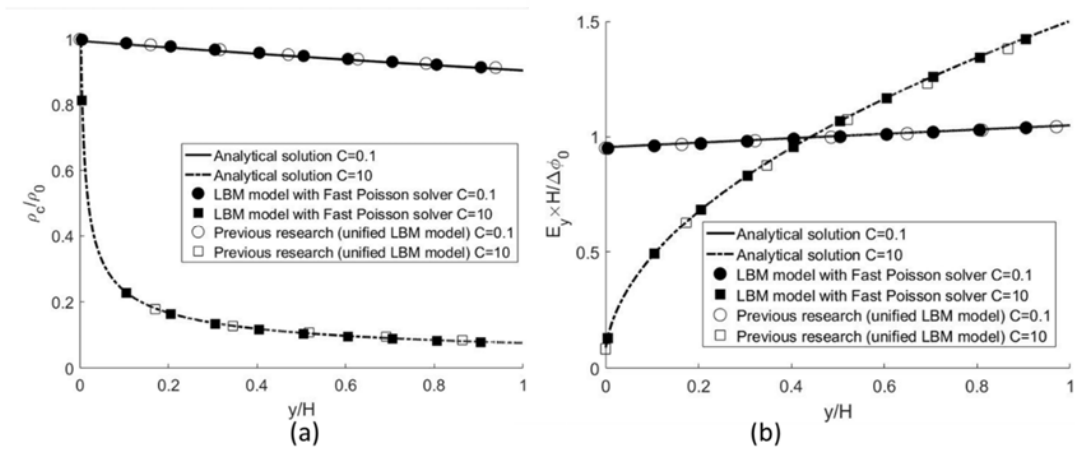

**FIG. 1.** Hydrostatic solution comparison of the TRT LBM and Fast Poisson solver, unified SRT LBM [18], and the analytical solution [30,40] for $C = 0.1$ and $C = 10$, $Fe = 4000$. (a) Electric field and (b) charge density;

Table II shows the dimensional parameters used for the analytical solution and the L$_2$ norm error between numerical results and analytical solutions. The numerical errors are lower than reported for the unified SRT LBM simulation ($e_{L_2}$ =0.0076) [18].

**Table II.** Dimensional parameters for the analytical model and L$_2$ norm errors $e_{L_2}$ for weak ion injection $C$ = 0.1 and strong ion injection $C$ = 10.

| $C$ | 0.1 | 10 |
|---|---|---|
| $\rho_a (Coulomb/m^{5/2})$ | 0.218 | 0.75 |
| $y_a(m)$ | 4.8 | 0.003 |
| $e_{L_2}$ of $E_y$ | 0.0031 | 0.0030 |
| $e_{L_2}$ of $\rho_c$ | 0.0035 | 0.0031 |

## 2. Electroconvection Instability

To model electro-convective instability, the steady-state hydrostatic solution is perturbed using waveform functions that satisfy the boundary conditions and continuity equation:

$$u_x = L_x \sin(2\pi y / L_y) \sin(2\pi x / L_x) \times 10^{-3}$$
$$u_y = L_y \left[ \cos(2\pi y / L_y) - 1 \right] \cos(2\pi x / L_x) \times 10^{-3} \quad (29)$$

The physical domain size is $Lx = 1.22m$ and $Ly = 1m$ limits the perturbation wavenumber to $\lambda_x = 2\pi / L_x \approx 5.15(1/m)$ -- the most unstable mode under the condition, where $C = 10, M = 10$, $Fe = 4000$ [19]. The electric Nusselt number $Ne_0 = I_1 / I_0$ acts as a criterion of flow stability, where $I_1$ is the cathode current at a specific condition, $I_0$ is the cathode current for the hydrostatic solution [5]. For cases where EC vortices exist, $Ne_0 > 1$. For a strong ion injection, the EC stability largely depends on $T$; so, in this analysis, $T$ is varied, while other parameters are held constant: $C = 10$, $M = 10$ and $Fe = 4000$.

For increasing $T$ ($T \geq T_c$) and the perturbation is given by Eq. 29, the flow becomes unstable developing EC vortices which are maintained by an electric force acting on the ionized fluid -- a combination of applied electric field and the space charge effect. The space charge effect can alter the applied electric field in the area of high charge density [41]. FIG. 2 (a) shows the formation of counter-rotating vortices; the charge density contour plotted with streamlines. In an upward fluid motion, the local charge transport is enhanced as indicated by the higher charge density in the center of the domain. In downward flow motion, the charge transport decreases, see the darker blue in the edges of the domain. FIG. 2 (b) shows the x-directional velocity contour. High x-velocity regions are located near the top and bottom walls; the flow is symmetric, which indicates that the steady-state solution has the same wavelength in x and y directions as the perturbation equations (Eq. 29).

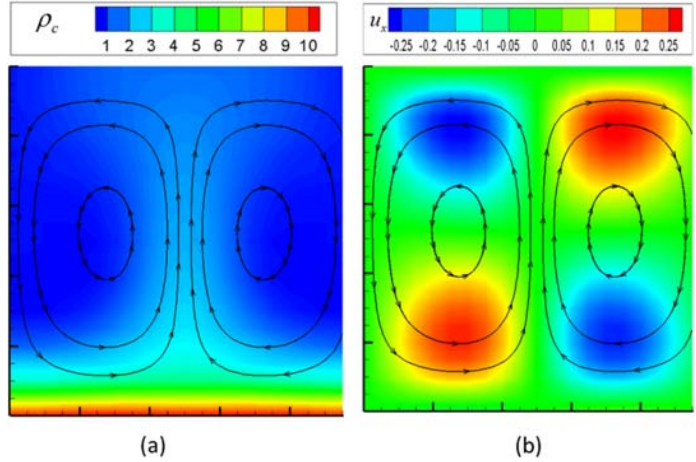

**FIG. 2.** Charge density and $u_x$ contours for EHD convection with vortices.

FIG. 3 (a) shows the EC flow stability analysis, demonstrated by $Ne_0$ as a function of $T$. When $T < T_c$ the perturbation does not trigger the flow instability, the perturbed flow reverses to the hydrostatic state. If $T$ decreases after the EC vortices are formed, they are maintained until $T = T_f$ when the system returns to the hydrostatic state. The model predicts the bifurcation points at $T_c = 163.4$ and $T_f = 108.7$ agreeing Luo et al. [19] ($T_c = 163.1$, $T_f = 108.7$), the linear stability analysis [12,19] ($T_c = 163.5$), and the finite volume method [16] ($T_f = 108.2$) under the same conditions. Neither numerical model or linear stability analysis agree with the experimental data. The proposed segregated TRT-LBM approach is consistent with the previous research; however, it does not modify governing equations by introducing artificial terms needed for numerical stability and yields fast convergence of the elliptical Poisson equation enabled for the FFT approach.

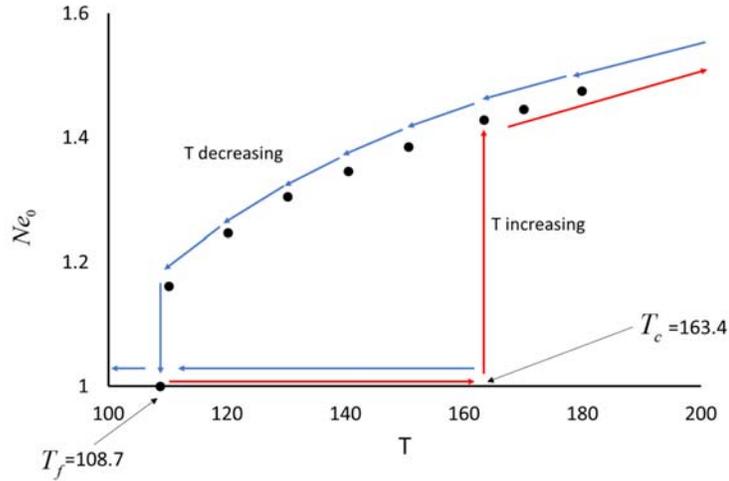

**FIG. 3.** Electric Nusselt number $Ne_0$ depends on electric Rayleigh number $T$

## VI. CONCLUSIONS

This work presents a numerical investigation of electroconvection phenomena between two parallel plates. The numerical approach combines (i) TRT-LBM for solving the transport equation of flow field and charged species, and (ii) Fast Poisson Solver. The TRT model allows for the use of two relaxation parameters, accounting for the difference between transport properties of neutral molecules and charged species. FFT algorithm for Poisson's equation directly solves for electric field enabling fast overall algorithm convergence. The numerical method is 2$^{nd}$ order accurate; it shows robust performance and agrees with previous results for the hydrostatic solution and for the solution where EC vortices are present.

# VII. ACKNOWLEDGMENTS

This research was supported by the DHS Science and Technology Directorate, Homeland Security Advanced Research Projects Agency, Explosives Division and UK Home Office, grant no. HSHQDC-15-531 C-B0033 and by the National Institutes of Health, grant NIBIB U01 EB021923.